\documentclass[aps,prl,twocolumn,amsmath,amsart,amssymb,superscriptaddress]{revtex4}
\usepackage{graphicx}
\usepackage[T1]{fontenc}
\usepackage{amsmath}
\usepackage{color}
\usepackage{times}

\usepackage{ulem}

\newcommand{\FSS}{FeSe$_{1-x}$S$_{x}$}

\begin{document}


\title{Nematic spin correlations pervading the phase diagram of {\FSS}}

\author{Ruixian Liu}
\affiliation{Center for Advanced Quantum Studies and Department of Physics, Beijing Normal University, Beijing, 100875 P. R. China}

\author{Wenliang Zhang}
\author{Yuan Wei}
\affiliation{Photon Science Division, Swiss Light Source, Paul Scherrer Institut, CH-5232 Villigen PSI, Switzerland}

\author{Zhen Tao}
\affiliation{Center for Advanced Quantum Studies and Department of Physics, Beijing Normal University, Beijing, 100875 P. R. China}
\affiliation{Photon Science Division, Swiss Light Source, Paul Scherrer Institut, CH-5232 Villigen PSI, Switzerland}

\author{Teguh C. Asmara}
\affiliation{Photon Science Division, Swiss Light Source, Paul Scherrer Institut, CH-5232 Villigen PSI, Switzerland}
\affiliation{European X-Ray Free-Electron Laser Facility GmbH, 22869 Schenefeld, Germany}

\author{Yi Li}
\affiliation{Center for Advanced Quantum Studies and Department of Physics, Beijing Normal University, Beijing, 100875 P. R. China}

\author{Vladimir N. Strocov}
\affiliation{Photon Science Division, Swiss Light Source, Paul Scherrer Institut, CH-5232 Villigen PSI, Switzerland}

\author{Rong Yu}
\affiliation{Department of Physics, Renmin University of China, Beijing 100872, China}

\author{Qimiao Si}
\affiliation{Department of Physics and Astronomy, Rice Center for Quantum Materials, Rice University, Houston, TX 77005, USA}

\author{Thorsten Schmitt}
\email{thorsten.schmitt@psi.ch}
\affiliation{Photon Science Division, Swiss Light Source, Paul Scherrer Institut, CH-5232 Villigen PSI, Switzerland}

\author{Xingye Lu}
\email{luxy@bnu.edu.cn}
\affiliation{Center for Advanced Quantum Studies and Department of Physics, Beijing Normal University, Beijing, 100875 P. R. China}

\date{\today}

\begin{abstract}

We use resonant inelastic X-ray scattering (RIXS) at the Fe-L$_3$ edge to study the spin excitations of uniaxial-strained and unstrained {\FSS} ($0\leq x\leq0.21$) samples. The measurements on unstrained samples reveal dispersive spin excitations in all doping levels, which show only minor doping dependence in energy dispersion, lifetime, and intensity, indicating that high-energy spin excitations are only marginally affected by sulfur doping. RIXS measurements on uniaxial-strained samples reveal that the high-energy spin-excitation anisotropy observed previously in FeSe is also present in the doping range $0< x\leq0.21$ of {\FSS}. The spin-excitation anisotropy persists to a high temperature up to $T>200$ K in $x=0.18$ and reaches a maximum around the nematic quantum critical doping ($x_c\approx0.17$).  Since the spin-excitation anisotropy directly reflects the existence of nematic spin correlations, our results indicate that high-energy nematic spin correlations pervade the regime of nematicity in the phase diagram and are enhanced by the nematic quantum criticality. These results emphasize the essential role of spin fluctuations in driving electronic nematicity and open the door for uniaxial strain tuning of spin excitations in quantum materials hosting strong magnetoelastic coupling and electronic nematicity.

\end{abstract}

\maketitle

Nematic order in quantum materials refers to an electronic state characterized by broken rotational symmetry and retained orientational order (or translational symmetry) \cite{Fradkin2015}. In iron-based superconductors (FeSCs), nematic order manifests as strong $C_2$ symmetric electronic anisotropies in the paramagnetic orthorhombic state with a small orthorhombic lattice distortion $\delta=(a-b)/(a+b)$ \cite{Fernandes2014, Si_NRM2016, Si_Hussey2023, Fernandes2022, Coldea2021, Chu2010, Chu2012, Yi2011, Lu2014, Kuo2016, Lu2018}. 
Under uniaxial strain ($\varepsilon$) along the $a/b$ axis, such anisotropy can persist to a temperature range beyond the nematic transition, revealing the fluctuating regime in the tetragonal state \cite{Chu2010, Chu2012, Yi2011, Lu2014, Kuo2016, Lu2018}. 
Nematic order/fluctuation has been demonstrated to be a universal feature of FeSCs \cite{Si_Hussey2023, Fernandes2022, Kuo2016}. 
Since electronic nematicity is essential for determining the exotic electronic properties of FeSCs and driving other emergent orders therein, it has attracted tremendous research interest in the past decade \cite{Fernandes2014,  Si_NRM2016, Si_Hussey2023, Fernandes2022, Coldea2021}.

{\FSS}, hosting exotic electronic properties and a pristine nematic transition without subsequent magnetic transition, has been a focus for studying electronic nematicity and its correlation with other intertwined order/fluctuations \cite{Si_Hussey2023, Coldea2018, Shibauchi2020, Coldea2021}. Figure 1(a) illustrates the compositional phase diagram of {\FSS} \cite{Coldea2021}. The parent compound FeSe exhibits a tetragonal-to-orthorhombic structural (nematic) transition at $T_s\approx90$ K and a superconducting transition at $T_c\approx8$ K \cite{Hsu2008, McQueen2009}. The nematic state at $T<T_s$ is a quantum-disordered magnetic state with intense antiferromagnetic (AF) spin excitations as revealed by neutron scattering and RIXS studies of twinned FeSe \cite{Wang2016_1, Wang2016_2, Rahn2019, Johnny21, FWang15, YuSi15}. Furthermore, our previous RIXS study on uniaxial-strain detwinned FeSe revealed underdamped spin excitations with strong anisotropy between the excitations along $[H, 0]$ and $[0, K]$ directions, which persists to a temperature slightly above $T_s$ under moderate uniaxial strain and has been suggested to be driven by spin nematicity \cite{Lu2022}. With increasing sulfur doping, the nematic transition is gradually suppressed to $T_s=0$ at the putative nematic quantum critical point (NQCP) ($x_c\approx0.17$), while $T_c$ increases slightly from $T_c\approx8$ K for $x=0$ to $T_c\approx10$ K at $x\approx0.08$ and decrease to $T_c\approx5$ K across the NQCP \cite{Coldea2021}. Beyond the nematic ordering region at $T\lesssim T_s$ and $x\lesssim x_c$, the nematic susceptibility of {\FSS} derived from elastoresistance measurements revealed a widely spread nematic fluctuating regime in the tetragonal phase ( $T\gtrsim T_s$ and $x\gtrsim x_c$)  \cite{Hosoi2016}.
The NQCP leads to maximized nematic susceptibility \cite{Hosoi2016}, and induces dramatic changes in various electronic properties \cite{Licciardello2019, Culo2021, Sato2018, Si_Hussey2023}. However, it is unclear how the AF spin excitations evolve across the nematic ordering phase boundary and the NQCP. In particular, on sulfur doping, as the nematic order weakens and disappears at the NQCP, it remains to be explored experimentally to what extent the spin-excitation anisotropy (nematic spin correlations) will pervade the phase diagram and be affected by the NQCP.

\begin{figure}
\includegraphics[width=8.5cm]{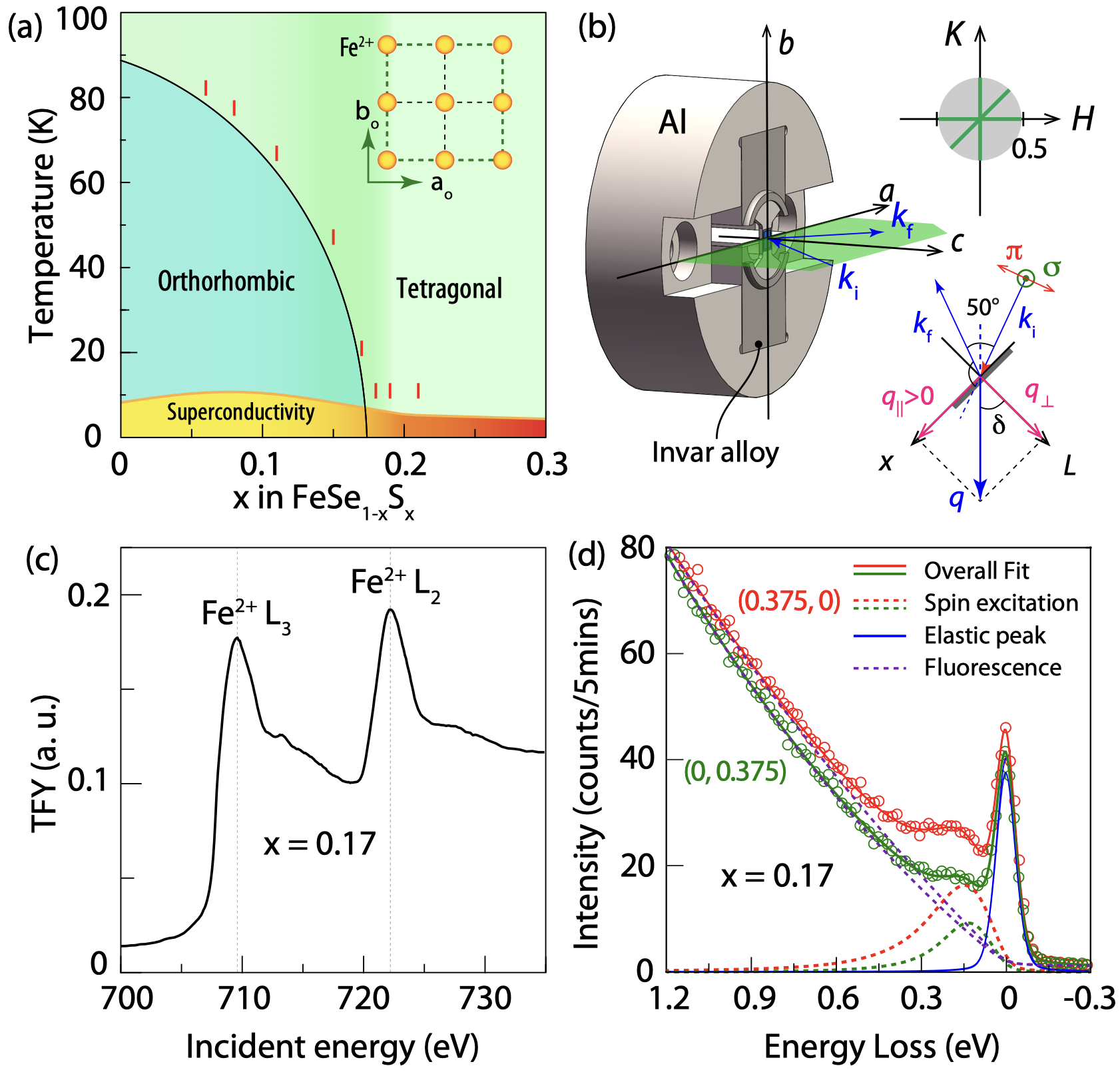}
\caption{ (a) Electronic phase diagram of {\FSS}. The red vertical bars mark the doping levels measured in the present study. (b) Schematics of scattering geometry with a crystal glued on a uniaxial strain device. The scattering angle was set to $2\theta_s=130^\circ$. (c) Total fluorescence yield (TFY) X-ray absorption spectroscopy (XAS) data of {\FSS} ($x=0.17$). (d) RIXS spectra of uniaxial-strained $x=0.17$ sample measured at $\mathbf{q}_\shortparallel = (0.375, 0)$ and $(0, 0.375), T=20$ K. The red and green solid curves are the overall fittings of the spectra. The red and green dashed curves are fittings of the spin excitations. The magenta dashed curves and the blue solid curves describe the fluorescence tail and the elastic peaks, respectively.  
}
\label{fig1}
\end{figure}

To address these issues, we use Fe-L$_3$ RIXS to measure the spin excitations of unstrained and uniaxial-strained {\FSS} ($x=0-0.21$) across the NQCP (Fig. 1(a)) \cite{rixs_rmp, chunjing_prx, kejin13, johnny19}. To explore the nematic spin correlations, one needs to apply uniaxial strain along $a/b$ axis of {\FSS} and measure the possible spin-excitation anisotropy \cite{Lu2022}. In this work, we use a uniaxial strain device based on differential thermal expansions of invar alloy and aluminum to apply uniaxial strain on {\FSS} (Fig. 1(b)) \cite{Sunko2019, SI}. While the measurements on unstrained samples reveal persistent spin excitations with minor doping dependence (Figs. 2 and 3), the measurements on uniaxial-strained samples reveal an enhancement of the spin-excitation anisotropy near the NQCP. Moreover, the prominent spin-excitation anisotropy persists in high doping levels ($x=0.21$) and temperature ($T>100$ K in $x=0.11$ and $T>200$ K in $x=0.18$) far beyond the nematic ordering region, demonstrating that the high-energy nematic spin correlations pervade a wide doping and temperature range in the phase diagram of {\FSS}.  Our results demonstrate that the electronic instability underlying the nematic fluctuating regime dominates a wide temperature and doping range in the phase diagram and significantly affects high-energy spin excitations. This discovery corroborates the spin-nematic picture and provides new insight for understanding the interplay between the intertwined orders/fluctuations in {\FSS}.

The doping levels studied in this work are marked by vertical red bars in the phase diagram of {\FSS} (Fig. 1(a)). 
Figure 1(b) illustrates the scattering geometry and the reciprocal space of the RIXS measurements, which were performed near the Fe-L$_3$ edge (incident energy $E_i\approx708$ eV) as shown in the total fluorescence yield (TFY) X-ray absorption spectroscopy (XAS) data in Fig. 1(c). The RIXS and XAS measurements were carried out with the RIXS experimental station at the ADRESS beamline of the Swiss Light Source at the Paul Scherrer Institut \cite{ADRESS, SAXES, SI}. 

To apply uniaxial strain on the sample, a thin {\FSS} crystal was glued onto the titanium bridge in the center of the uniaxial-strain device using epoxy (Fig. 1(b)). Upon cooling, the differential thermal expansion coefficients between the aluminum frame ($\alpha \approx -24\times10^{-6}$/K) and the invar-alloy blocks ($\alpha\approx-2\times10^{-6}$/K) can generate a uniaxial strain up to $\varepsilon=\varepsilon_{xx}-\varepsilon_{yy}\approx-0.8\%$ on the neck of the titanium bridge at low temperature, which can be transferred to the thin crystal glued on it \cite{Sunko2019, SI}. The uniaxial strain on the sample's surface can be accurately measured using a microscope \cite{SI}.


\begin{figure*}[htbp!]
\includegraphics[width=17 cm]{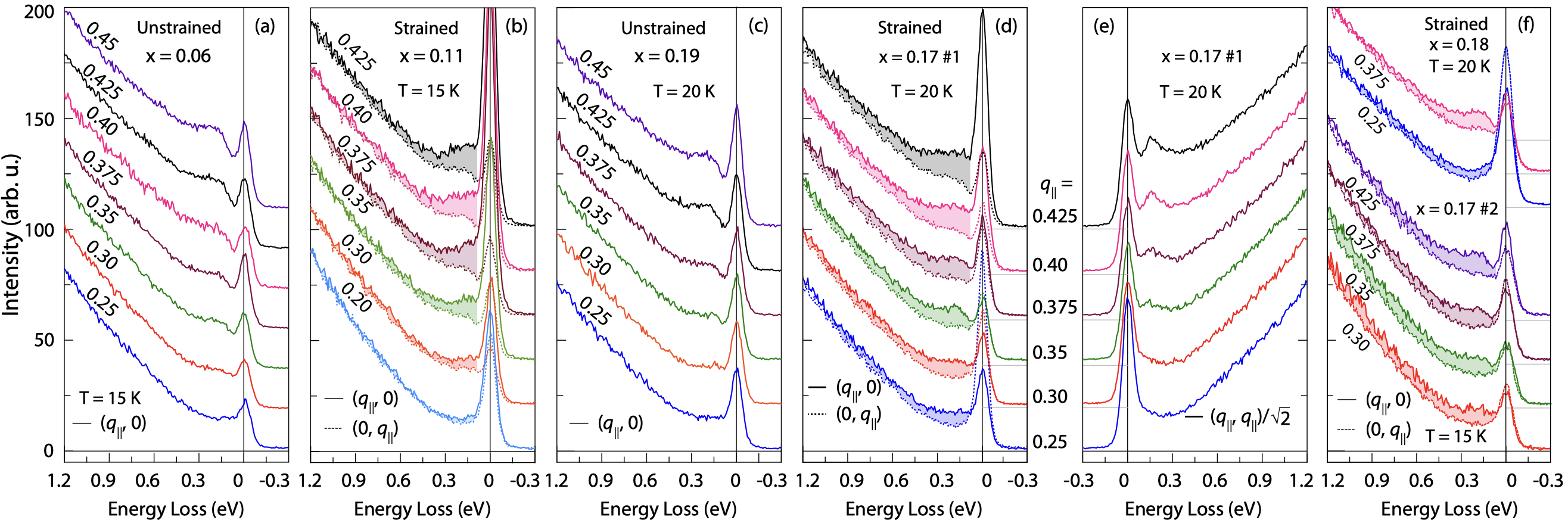}
\caption{
(a), (c) $q_\shortparallel$-dependent RIXS spectra along $H$ direction of unstrained $x=0.06$ and $0.19$ samples. (b) RIXS spectra along $H$ (solid curves) and $K$ (dashed curves) directions of uniaxial-strained $x=0.11$ sample. (d), (e) RIXS spectra measured along (d) $H/K$ and (e) $[H, H]$ directions of uniaxial-strained $x=0.17$ sample \#1. (f) RIXS spectra along $H/K$ directions of uniaxial-strained $x=0.17$ (sample \#2) and $0.18$. The color areas mark the difference between $I_h(q_\shortparallel)$ and $I_k(q_\shortparallel)$.}
\label{fig2}
\end{figure*}

To facilitate the discussion of the results presented below, we denote the RIXS spectra and the spin excitations along $H, K$ and $[H, H]$ directions as $I_{h/k/hh}(q_\shortparallel, E)$ and $S_{h/k/hh}(q_\shortparallel, E)$, respectively. The RIXS spectrum collected at $\mathbf{q}_\shortparallel$ can be expressed as
\begin{equation}
I(E) = I_{\rm fluo}(E) + S(E) + I_{\rm el}(E),
\label{Eq1}
\end{equation}
where $I_{\rm fluo}(E)$, $I_{\rm el}(E)$, and $S(E)$ are the intensity vs. energy loss of fluorescence, elastic peak, and spin excitation, respectively. 
Figure 1(d) shows two representative RIXS spectra $I_h(q_\shortparallel, E)$ (red circles) and $I_k(q_\shortparallel, E)$ (green circles) ($q_\shortparallel=0.375$) of an $x=0.17$ sample measured under a uniaxial strain of $\varepsilon\approx-0.6\%$ (at $T=20$ K) along the $b$ axis. Because the spin excitations of twined/unstrained FeSC crystals would be $C_4$ symmetric ($I_h(q_\shortparallel, E)=I_k(q_\shortparallel, E))$ in the $[H, K]$ reciprocal space, the substantial difference between $I_h(q_\shortparallel, E)$ and $I_k(q_\shortparallel, E)$ clearly reveals an excitation anisotropy with $C_2$ symmetry induced by the uniaxial strain. 
The fitting analysis of the spectra following Eq. (1) agrees well with the raw data (red and green circles) \cite{SI} and reveals a strong anisotropy between  $S_h(q_\shortparallel, E)$ (red dashed curve) and $S_k(q_\shortparallel, E)$ (green dashed curve), unveiling the existence of nematic spin correlations in {\FSS}. 

\begin{figure}[htbp!]
\includegraphics[width=8cm]{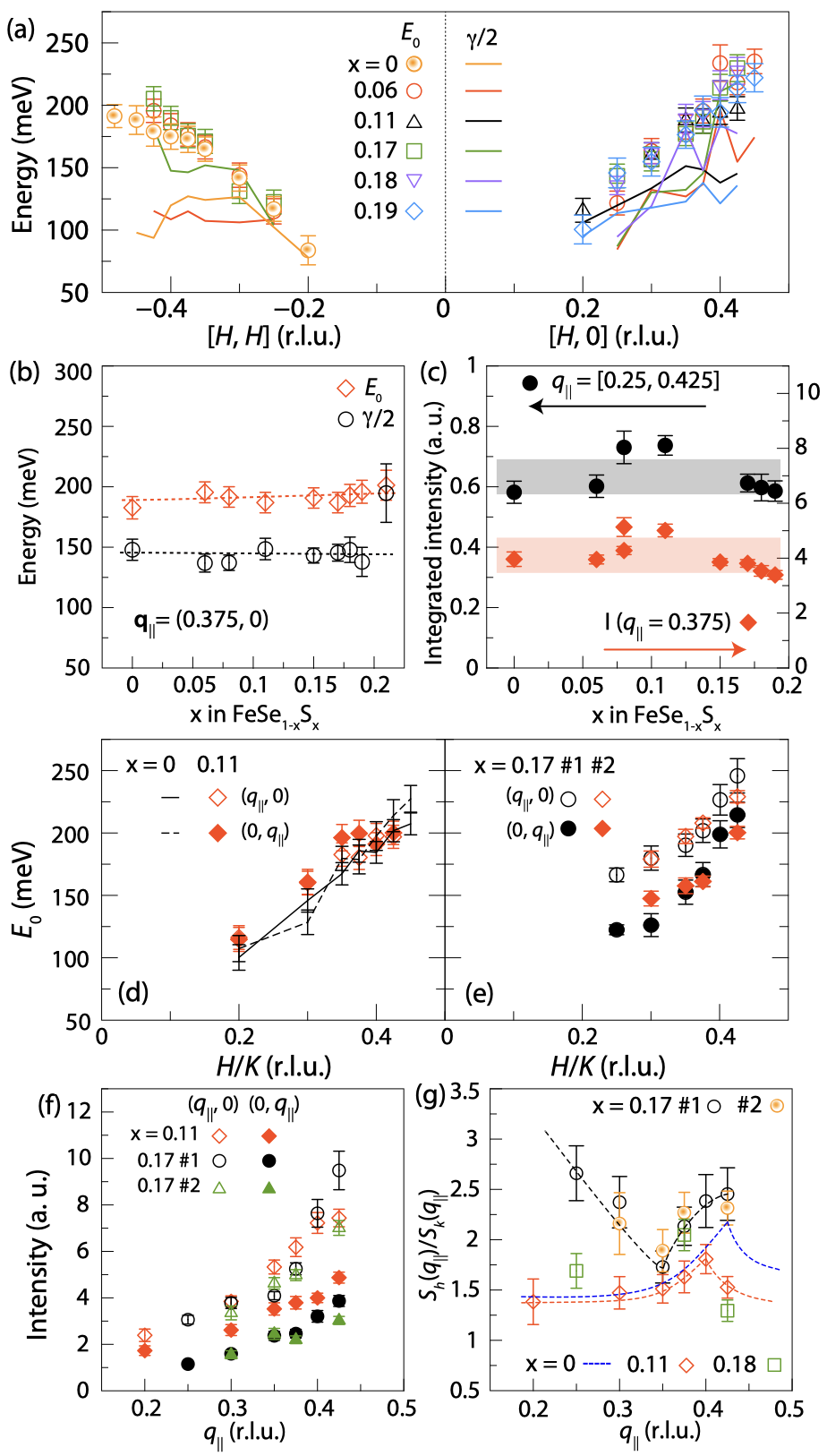}
\caption{
(a) $E_0$ and $\gamma/2$ of the spin excitations along $H$ ($S_h(q_\shortparallel, E))$ and $[H, H]$ ($S_{hh}(q_\shortparallel, E))$ for unstrained $x=0.06, 0.11, 0.17, 0.18$, and $0.19$. (b) $E_0$ and $\gamma/2$ of the spin excitation at $H=0.375$ of unstrained samples as a function of the doping level $x$. (c) Doping dependence of the spin-excitation intensity $S_h(q_\shortparallel)$ for $q_\shortparallel=0.375$ (red diamonds) and the momentum integrated $S_h(q_\shortparallel)$ in the momentum range $q_\shortparallel=[0.25, 0.425]$ (black circles). (d), (e) Comparison of  $E_0(q_\shortparallel)$ between $S_h(q_\shortparallel, E)$ and $S_k(q_\shortparallel, E)$ for (d) $x=0$ (black lines), $0.11$ (red diamonds) and (e) $x=0.17$. (f) Energy integrated spin-excitation intensity $S_h(q_\shortparallel)$ (open symbols) and $S_k(q_\shortparallel)$ (filled symbols) for $x=0.11$ (red diamonds), $0.17$ \#1 (black circles), and $0.17$ \#2 (green triangles). (g) Doping-dependent spin-excitation anisotropy $S_h(q_\shortparallel)/S_k(q_\shortparallel)$ for $x=0$ (blue dashed curve), $0.11$ (red diamonds), $0.17$ (black and illuminated orange circles), and $0.18$ (green squares). The red and black dashed curves are guides to the eyes.}
\label{fig3}
\end{figure}

Having established the experimental strategy, we have carried out systematic RIXS measurements on unstrained and uniaxial-strained {\FSS} samples in the doping range $x=0.06 - 0.21$ to sort out the development of the spin excitations and their anisotropy (Figs. 2-4) across the electronic phase diagram. Figure 2 displays the momentum-dependent RIXS spectra measured at $T=15$ and $20$ K on unstrained $x=0.06$ (Fig. 2(a)), strained $x=0.11$ ($\varepsilon\approx-0.8\%$ Fig. 2(b)), unstrained $x=0.19$ (Fig. 2(c)), strained $x=0.17$ \#1 ($\varepsilon\approx-0.6\%$, Figs. 2(d), 2(e)), and strained $x=0.18$ ($\varepsilon\approx-0.4\%$, Fig. 2(f))  and the $x=0.17$ \#2 ($\varepsilon\approx-0.6\%$, Fig. 2(f)) samples, in which $x=0.06$ and $0.11$ are in the nematic ordering regime, $x=0.17$ at the NQCP, and $x=0.19$ in the tetragonal phase \cite{SI}. All the spectra were normalized according to the doping-independent fluorescence in the energy-loss range of [1, 10] eV.
The waterfall plots in Fig. 2 show that dispersive spin excitations along $H/K$ and $[H, H]$ directions are observed in a wide doping range covering the nematic regime. For the unstrained samples, the momentum-dependent RIXS spectra of $x=0.19$ are similar to those of the $x=0.06$ ($T_s\approx74$ K) sample, implying only subtle change of the spin excitations across the phase boundary of the nematic order. For the strained samples, the RIXS spectra collected both in the nematic ordering regime ($x=0.11$, $T_s\approx65$ K), and close to the NQCP ($x=0.17$ and $0.18$ samples), exhibit prominent excitation anisotropy ($I_h(q_\shortparallel, E)$>$I_k(q_\shortparallel, E)$ in Fig. 2.

To extract the information about the spin excitations and understand the spin-excitation anisotropy quantitatively, we use a damped harmonic oscillator function  \cite{Rahn2019, Lu2022} to describe the spin excitations collected with RIXS:
\begin{equation}
S(q,E)=A\,\frac{2\,\gamma\,E}{\left(E^2-E_0^2\right)^2+(\gamma\,E)^2},
\label{Eq1}
\end{equation}
where $E_0(\mathbf{q}_\shortparallel)$ is the undamped energy and $\gamma(\mathbf{q}_\shortparallel)$ is the damping factor. 
Generally, the undamped energy $E_0(\mathbf{q}_\shortparallel)$ and the excitation lifetime $\gamma(\mathbf{q}_\shortparallel)$ are essential information to characterize line-shape of the excitations. If in the fitting with Eq (2) $E_0>\gamma/2$ ($E_0<\gamma/2$), the spin excitations are defined as underdamped (overdamped).  

Figure 3 summarizes the fitting results of the RIXS spectra collected in the doping range covering the nematic phase and the NQCP ($x_c\approx0.17$). Figure 3(a) shows the energy dispersion $E_0(\mathbf{q_\shortparallel})$ (colored symbols) and the damping rate $\gamma/2$ (colored lines) for the spin excitations of unstrained samples, for which the unstrained $I_h(q_\shortparallel, E)$ of $x=0.11$ and $0.17$ were obtained by averaging the $I_h(q_\shortparallel, E)$ and $I_k(q_\shortparallel, E)$ of uniaxial-strained samples. The energy dispersion $E_0(\mathbf{q_\shortparallel})$ exhibits negligible doping dependence along both $[H, 0]$ and $[H, H]$ directions. Moreover, the damping rate $\gamma/2$ also shows a minor doping dependence. As $\gamma/2$ is overall lower than $E_0$ in the whole momentum range studied, all the spin excitations observed here are underdamped paramagnons. The doping dependence of $E_0$ and $\gamma/2$ indicates that the high-energy spin excitations are not affected by sulfur doping. This is further evidenced by the doping-independent $E_0$ and $\gamma/2$ (Fig.3(b)), and the largely unchanged energy-integrated intensity $S_h(q_\shortparallel)$ (red diamonds in Fig. 3(c)) for $\mathbf{q_\shortparallel}=(0.375, 0)$ in more dopings. For $S_h(q_\shortparallel=0.375)$, $E_0$ is constantly larger than $\gamma/2$, consistent with underdamped spin excitation modes. 
Figure 3(c) shows the energy- and $q_\shortparallel$-integrated (within the momentum integral interval ${q_\shortparallel}=[0.25, 0.425]$) intensity of the spin excitations along the $H$ axis of unstrained samples (black filled circles). Again, the integrated spectral weight does not show a clear increasing or decreasing tendency, i.e., it remains unchanged from $x=0$ to $0.19$.

\begin{figure}[htbp!]
\includegraphics[width=8.5 cm]{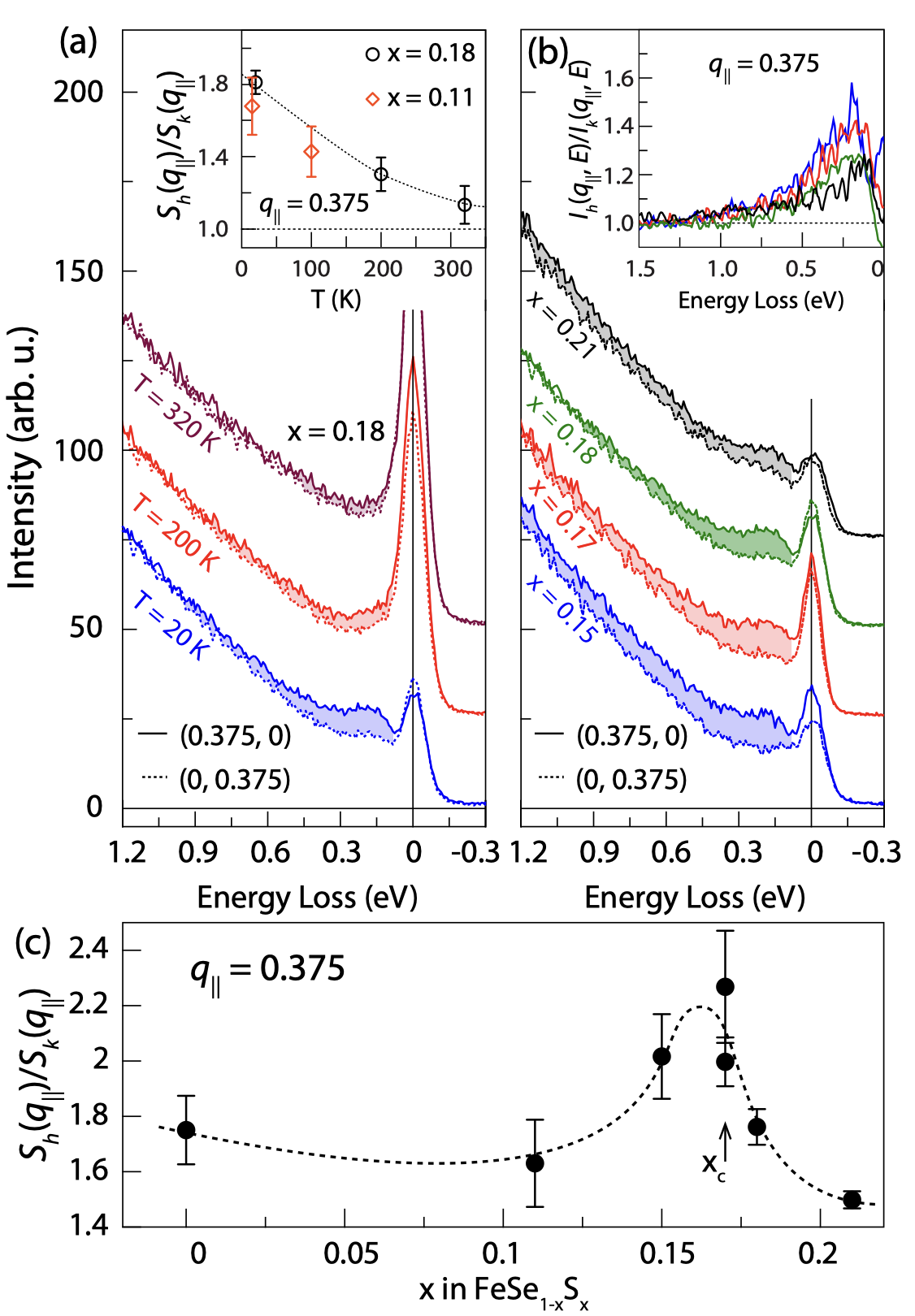}
\caption{Temperature and doping dependence of the spin-excitation anisotropy of uniaxial-strained {\FSS}. (a) Temperature dependence of the RIXS spectra for $\mathbf{q}_{\shortparallel} = (0.375, 0)$ and $(0, 0.375)$ for $x=0.18$ measured at $T = 20, 200$ and $320$ K with $\varepsilon\approx-0.6\%$ at $T = 20$ K. The inset shows the temperature-dependent intensity ratio $S_h(q_\shortparallel)/S_k(q_\shortparallel)$ with $q_\shortparallel=0.375$ for $x=0.11$ and $0.18$. $S_h(q_\shortparallel)/S_k(q_\shortparallel)$ are obtained from the fitting of the spin excitations. (b) Doping-dependent RIXS spectra ($x = 0.15, 0.17, 0.18$ and $0.21$) with $\mathbf{q}_
\shortparallel = (0.375, 0)$ and $(0, 0.375)$. The colored areas mark the difference between the spectra $I_h(q_{\shortparallel}$) and $I_k(q_{\shortparallel}$). The inset shows the ratio between $I_h(q_{\shortparallel}$) and $I_k(q_{\shortparallel})$ for $x=0.15, 0.17 (\#1), 0.18$ and $0.21$. (c) Doping dependence of the $S_h(q_\shortparallel)/S_k(q_\shortparallel)$ with $q_\shortparallel=0.375$.}

\label{fig4}
\end{figure}


Figure 3(d) exhibits the undamped energy dispersion $E_0(\mathbf{q_\shortparallel})$ of $S_h(q_\shortparallel, E)$ and $S_k(q_\shortparallel, E)$ for uniaxial-strain detwinned $x=0$ and $0.11$ ($T_s\approx65$ K).  The $E_0(\mathbf{q_\shortparallel})$ along $H$ and $K$ directions overlap in these two dopings. In comparison, $S_h(q_\shortparallel, E)$ and $S_k(q_\shortparallel, E)$ in $x=0.17$ exhibit a clear difference in $E_0(\mathbf{q_\shortparallel})$ (Fig. 3(e)), indicating that the uniaxial strain can induce a pronounced difference in the energy dispersion of spin excitations near the NQCP. Figure 3(f) displays the energy integrated intensity of the spin excitations $S_h(q_\shortparallel)$ and $S_k(q_\shortparallel)$ for $x=0.11$, and $0.17$, from which we determine the ratio $S_h(q_\shortparallel)/S_k(q_\shortparallel)$ (Fig. 3(g)), which reflects the magnitude of spin-excitation anisotropy. The spin-excitation anisotropy in $x=0.11$ is overall slightly lower than that of FeSe (blue dashed curve), indicating that the intrinsic nematic spin correlations in the nematic state weaken as the sulfur doping increases, which suppresses the nematic transition.

With further increase of doping, the nematic order is completely suppressed at $x=0.17$; i.e., this doping creates a tetragonal system at finite temperature. Surprisingly, RIXS measurements on $x=0.17$ with $\varepsilon\approx-0.6\%$ reveal a pronounced spin-excitation anisotropy that is even stronger than those in detwinned FeSe ($\varepsilon\approx-2\delta\approx-0.54\%$ at $T=20$ K) and FeSe$_{0.89}$S$_{0.11}$ ($\varepsilon\approx-0.8\%$ at $T=15$ K)  (Fig. 3(g)). 
Since the $x=0.17$ sample falls in the regime of nematic quantum criticality,  the enhancement of the nematic spin correlations could be driven by strong nematic fluctuations near the NQCP \cite{fnote}.

 To further assess the enhancement of nematic spin correlations near the NQCP and illustrate the nematic spin correlations in the fluctuating regime, 
 we show in Fig. 4 the temperature and doping dependence of the spin-excitation anisotropy $S_h(q_\shortparallel)/S_k(q_\shortparallel)$ at $q_\shortparallel=0.375$. We find that the spin-excitation anisotropy $S_h(q_\shortparallel)/S_k(q_\shortparallel)\approx1.65$ at $T=15$ K in $x=0.11$ decreases only by $\sim15\%$ when warmed to $T=100$ K ($>T_s=65$ K) (inset of Fig. 4(a)), indicating that the nematic spin correlations can persist to higher temperatures well above the $T_s$ in the nematic ordering regime ($x<x_c$). Furthermore, we show in Fig. 4(a) $I_h(q_\shortparallel, E)$ and $I_k(q_\shortparallel, E)$ of $x=0.18$ (slightly higher than $x_c$) measured under a moderate uniaxial strain $\varepsilon\approx-0.4\%$ ($T=20$ K). The spectra show clear difference at $T=20$ K, generating $S_h(q_\shortparallel)/S_k(q_\shortparallel) \approx1.8$ (inset of Fig. 4(a)). Upon warming, while the uniaxial strain was relaxed gradually, the anisotropy, however, persists at $T=200$ K (decreased by $\sim40\%$ ), and finally vanishes at $T=320$ K (Fig. 4(a)). 
 
Figure 4(b) shows the spectra $I_h(q_\shortparallel, E)$ and $I_k(q_\shortparallel, E)$ measured at the same $q_\shortparallel=0.375$ near the regime of NQC. Similar spectral difference is observed in $x=0.15$ ($T_s=50$ K, $\varepsilon \approx -0.6\%$) and $x=0.21$ $(\varepsilon\approx-0.4\%)$ at $T=20$ K (color areas in Fig. 4(b)). The inset of Fig. 4(b) depicts the doping dependence of the spin excitation anisotropy $I_h(q_\shortparallel, E)/I_k(q_\shortparallel, E)$ for $x=0.15$ ($\varepsilon\approx-0.6\%$, blue), $x=0.17$ ($\varepsilon\approx-0.6\%$, red), $x=0.18$ ($\varepsilon\approx-0.4\%$, green), and $x=0.21$ ($\varepsilon\approx-0.4\%$, black), revealing a decreasing tendency for higher doping. We note that the anisotropy of the RIXS spectra persists to an energy scale $\sim1$ eV of the electron-hole pair tail, which could be attributed to an anisotropy in incoherent charge scattering \cite{Degiorgi2017}. 

Figure 4(c) summarizes the doping-dependent spin excitation anisotropy $S_h(q_\shortparallel)/S_k(q_\shortparallel)$ with $q_\shortparallel=0.375$ extracted from the fitting of the spin excitations \cite{SI}. Consistent with the enhancement at $x=0.17$ (Fig. 3(g)), the anisotropy in Fig. 4(c) reaches a maximum in the doping region $x\approx0.15-0.18$.  Such a doping evolution of the nematic spin fluctuations is in line with the exotic electronic properties at the NQCP ($x_c\approx0.17$) \cite{Hosoi2016, Licciardello2019, Culo2021, Sato2018}. It is well known that quantum critical fluctuations usually dominate static electronic transport properties and low-energy charge/spin dynamics \cite{Sachdev_book}. The $E \sim 200$ meV nematic spin correlations exhibiting a maximum near the NQCP strongly suggest that the nematic fluctuations can also dominate the physics at a much higher energy scale (under a uniaxial-strain field).

The nematic fluctuating regime in FeSCs has been well established in various studies of (quasi-)static properties, such as the softening of the shear modulus $C_{66}$ \cite{Anna2014, Anna2015, Fujii2018}, the divergent nematic susceptibility obtained from elastoresistance $-2m_{66}$ \cite{Kuo2016, Hosoi2016}, and the persistence of local orthorhombicity (short-range orthorhombic structure) in the tetragonal state of both iron pnictides and FeSe \cite{Lu2016, Ben2017, Weiyi18, Ben2019}. 
As such local orthorhombicity persists in both iron pnictides and iron chalcogenides, it should be a common feature of the nematic fluctuating regime of FeSCs. Our identification of spin-excitation anisotropy at high energies is consistent with the general notion that high-energy fluctuations are concomitant with short-spatial-range correlations. Our results, thus, uncover a common feature that high-energy nematic spin fluctuations permeate across FeSCs.

In summary, we find the high-energy spin-excitations in {\FSS} ($x\lesssim0.21$) are only marginally affected by sulfur doping, and illustrate that the nematic spin correlations pervade wide doping ($0\lesssim x\lesssim 0.21$) and temperature region of the phase diagram, corroborating the spin-nematic picture.
The enhancement of the nematic spin correlations near the NQCP suggests that the nematic fluctuations could affect the physics at a higher energy scale. The strain-induced spin-excitation anisotropy in the tetragonal state of {\FSS} establishes uniaxial strain as an effective way to tune spin and/or charge fluctuations in similar quantum materials hosting electron-lattice coupling.

\begin{acknowledgments}
The work at Beijing Normal University is supported by the National Key Projects for Research and Development of China (Grant No. 2021YFA1400400) and the National Natural Science Foundation of China (Grant Nos. 12174029, and 11922402) (X.L.). The RIXS experiments were carried out at the ADRESS beamline of the Swiss Light Source at the Paul Scherrer Institut (PSI). The work at PSI is supported by the Swiss National Science Foundation through project no. 200021\_207904, 200021\_178867, and the Sinergia network Mott Physics Beyond the Heisenberg Model (MPBH) (project numbers CRSII2 160765/1 and CRSII2 141962). Y.W. and T.C.A. acknowledge financial support from the European Union’s Horizon 2020 research and innovation programme under the Marie Skłodowska-Curie grant agreement No. 884104 (Y.W.) and No. 701647 (T.C.A.) (PSI-FELLOW-III-3i) 
Work at Renmin has in part been supported by the National Science Foundation of China Grant No. 12174441. 
Work at Rice was primarily supported by the U.S. Department of Energy, Office of Science, Basic Energy Sciences, under Award No. DE-SC0018197, 
and by the Robert A. Welch Foundation Grant No. C-1411. Q.S. acknowledges the hospitality of the Kavli Institute for Theoretical Physics, supported in part by the National Science Foundation under Grant No. NSF PHY1748958, during the program ``A Quantum Universe in a Crystal: Symmetry and Topology across the Correlation Spectrum", as well as the hospitality of the Aspen Center for Physics, which is supported by NSF grant No. PHY-1607611.
\end{acknowledgments}



\begin{thebibliography}{}
\bibitem{fnote} It is worth noting that these unusual discoveries on strained $x=0.17$ was first observed in sample \#1, and confirmed in sample \#2 in a separate RIXS experiment.
\end{thebibliography}
\end{document}